# DOCTORS: a computational framework for photon transport simulation in X -ray radiography and tomography based on the discrete ordinates method

Edward T. Norris, PhD
Steve Wagstaff
Xin Liu, PhD*

Missouri University of Science and Technology,
Nuclear Engineering and Science, Rolla, MO

*Corresponding author
301 W 14th Street
Rolla, MO 65401
Phone: (573) 341-4693
Fax: (573) 341-4720
xinliu@mst.edu

**Abstract**

Accurately simulating photon transport is crucial for non-destructive testing and medical diagnostic applications using X-ray radiography and computed tomography (CT). Solving a discretized form of the linear Boltzmann equation is a computationally efficient technique for simulating photon transport. We have developed a software package, called "DOCTORS", which solves the linear Boltzmann equation using the discrete ordinates method. The computational framework and its GPU implementation are described in detail. The accuracy of DOCTORS simulations was verified by quantitative comparisons to Monte Carlo simulations. Photon fluence distributions, as calculated by DOCTORS, agreed well with Monte Carlo simulations, with near real-time computation speed.

**Keywords:** photon transport, Boltzmann equation, discrete ordinates, radiography, computed tomography, GPU

## I. Introduction

Low energy ($E$ < 1 MeV) X-ray imaging, including radiography and CT, is an important non-destructive testing and evaluation technique in numerous industrial and medical applications. A complete description of the photon distribution and energy transfer is essential for estimating radiation doses and designing an optimized imaging system. Stochastic methods (e.g., Monte Carlo simulation) have been used extensively in the past, [1-6] and are generally considered the gold standard for estimating photon distributions and radiation doses. However, stochastic methods require a large number of particle histories and, therefore, need a lengthy computation time to reduce statistical uncertainties to acceptable levels. However, no statistical error is associated with deterministic methods, so they can be comparatively efficient in regions where the highly resolved spatial fluence must be known within a tight uncertainty bound. To date, we have developed several deterministic simulations of a CT system and its subcomponents.[7-11] In the course of these efforts, we discovered that the deterministic solution of the linear Boltzmann equation, based on the discrete ordinates method (i.e., $S_N$ method), is the most promising method, due to its scalability and parallelizability. Computer codes, based on the discrete ordinates method, have been extensively used in radiation shielding calculations and nuclear reactor analyses. They have not, however, been applied to low energy X-ray radiography and CT. Recently, we have developed a software application called DOCTORS (Discrete Ordinates Computed TOmography and Radiography Simulator).[12] In this paper, we describe the computational framework and demonstrate its accuracy by computing energy-resolved photon fluence distributions in both homogenous and inhomogeneous phantoms resulting from cone-beam X-ray radiography and tomography. Precompiled DOCTORS will be freely distributed to the public.

## II. Methods

### II.A. Deterministic Photon Transport Model and Discrete Ordinates Method

The photon transport process can be described by the steady-state linear Boltzmann transport equation. The steady-state linear Boltzmann transport equation is given by the following:[13]

$$\left[\hat{\Omega} \cdot \vec{\nabla} + \sigma_t(\vec{r}, E)\right]\phi(\vec{r}, E, \hat{\Omega}) = \int_0^\infty \int_{4\pi} \sigma_s(\vec{r}, E' \rightarrow E, \hat{\Omega} \cdot \hat{\Omega}')\phi(\vec{r}, E', \hat{\Omega}')d\hat{\Omega}'dE' + S(\vec{r}, E, \hat{\Omega}) \tag{1}$$

where, $\phi(\vec{r}, E, \hat{\Omega})$ is the angular fluence (photons/cm$^2$) at position $\vec{r}$, with energy $E$, and direction $\hat{\Omega}$; $\sigma_t(\vec{r}, E)$ is the total macroscopic interaction cross section, including scattering and absorption cross sections; $S(\vec{r}, E, \hat{\Omega})$ is the external source which simplifies the actual physical source; $\sigma_s(\vec{r}, E' \rightarrow E, \hat{\Omega} \cdot \hat{\Omega}')$ is the macroscopic differential scattering cross section which represents the probability that photons at position $\vec{r}$ are scattered from energy $E'$ and direction $\hat{\Omega}'$ to energy $E$ and direction $\hat{\Omega}$; $\hat{\Omega} \cdot \hat{\Omega}'$ is the cosine of the scattering angle. The left side of the Equation (1) represents photon loss in a differential volume at position $\vec{r}$, and the right side represents the gain of photons in the same differential volume and position.

Analytical solutions of the Boltzmann transport equation can only be obtained for the simplest problems. Realistic, multidimensional, and energy-dependent problems must be solved numerically. The discrete ordinates method discretizes the continuous variables, $\vec{r}$, $E$, and $\hat{\Omega}$, onto a discrete phase space so that Equation (1) can be solved numerically.[13] Detailed derivation of discretized variables can be found in many publications.[10, 13] Only the final form of a discretized linear Boltzmann transport equation is given here,

$$\left[ \hat{\Omega}_n \cdot \vec{\nabla} + \sigma_{i,j,k}^g \right] \phi_{i,j,k,n}^g = \sum_{n=1}^{N} \sum_{g'=g}^{G-1} \Sigma_{s,nn'}^{gg'} \phi_{i,j,k,n'}^{g'} \omega_n + S_{i,j,k,n}^g \tag{2}$$

where, $n$ represents the discrete direction; $i$, $j$, and $k$ represent the 3D spatial discrete mesh grid; $g$ represents the energy group. The angular fluence $\phi_{i,j,k,n}^g$ in equation (2) can be solved numerically with proper boundary conditions. In our study, vacuum boundary conditions were applied. The scalar fluence at each voxel ($i$, $j$, $k$) was obtained by integration over all of the discrete angles and approximated by the quadrature formula,

$$\phi_{i,j,k}^g = \sum_{n=1}^{N(N+2)} w_n \phi_{i,j,k,n}^g \tag{3}$$

where $N$ is the discrete ordinates order that are referred to as $S_N$ quadratures, which resulted in a total of $N(N+2)$ directions; $w_n$ is the weight associated with each discrete direction.

Discrete ordinates methods suffer from 'ray-effect' in a weakly scattered media.[13] A 'ray-effect' arises due to the restriction of particle transport to a set of discrete directions. To mitigate this 'ray-effect', the first-collision source method was employed.[13] The fluence in each voxel was composed of uncollided photons directly from the source and collided photons scattered from other voxels, energies, and directions. Thus, equation (1) can be rewritten as two equations for solving the uncollided fluence $\phi_u$, and collided fluence, $\phi_c$ independently. Both $\phi_u$ and $\phi_c$ obey the linear Boltzmann transport equation and the sum of $\phi_u$ and $\phi_c$ is $\phi$. In equation (4), the uncollided photon transport equation has no scatter term since scattered particles were not considered in the uncollided fluence. The lack of a scatter term makes the uncollided fluence computable with a raytracing algorithm. The uncollided fluence is then used to compute the first-collision source, $S_u$, as shown in equation (5). The first-collision source is used to drive the collided fluence in the same way that the external source drives the uncollided fluence distribution as shown in equation (6). Using equation (5) to substitute $S_u$ in equation (6), simplifies equation (6) to equation (7), which can be solved iteratively by using the discrete ordinates method described above.

$$[\hat{\Omega} \cdot \vec{\nabla} + \sigma_t(\vec{r}, E)]\phi_u(\vec{r}, E, \hat{\Omega}) = S(\vec{r}, E, \hat{\Omega}) \tag{4}$$

$$S_u(\vec{r}, E, \hat{\Omega}) = \int_0^\infty \int_{4\pi} \sigma_S(\vec{r}, E' \rightarrow E, \hat{\Omega} \cdot \hat{\Omega}')\phi_u(\vec{r}, E', \hat{\Omega}')d\hat{\Omega}'dE' \tag{5}$$

$$\left[\hat{\Omega} \cdot \vec{\nabla} + \sigma_t(\vec{r}, E)\right]\phi_c(\vec{r}, E, \hat{\Omega}) = \int_0^\infty \int_{4\pi} \sigma_s(\vec{r}, E' \rightarrow E, \hat{\Omega} \cdot \hat{\Omega}')\phi_c(\vec{r}, E', \hat{\Omega}')d\hat{\Omega}'dE' + S_u(\vec{r}, E, \hat{\Omega}) \tag{6}$$

$$[\hat{\Omega} \cdot \vec{\nabla} + \sigma_t(\vec{r}, E)]\phi_c(\vec{r}, E, \hat{\Omega}) = \int_0^\infty \int_{4\pi} \sigma_S(\vec{r}, E' \rightarrow E, \hat{\Omega} \cdot \hat{\Omega}')\phi_c(\vec{r}, E', \hat{\Omega}')d\hat{\Omega}'dE' +$$
$$\int_0^\infty \int_{4\pi} \sigma_S(\vec{r}, E' \rightarrow E, \hat{\Omega} \cdot \hat{\Omega}')\phi_u(\vec{r}, E', \hat{\Omega}')d\hat{\Omega}'dE' \tag{7}$$
$$= \int_0^\infty \int_{4\pi} \sigma_S(\vec{r}, E' \rightarrow E, \hat{\Omega} \cdot \hat{\Omega}')\phi(\vec{r}, E', \hat{\Omega}')d\hat{\Omega}'dE'$$

## II.B. Accelerate Computation through GPU Parallel Computing

For simulations of large systems, the computation costs are daunting due to a large number of discrete angles, energy groups, and voxels. To accelerate the computation process, GPU computing techniques were adopted in DOCTORS.

The GPU parallelization of the ray-tracing algorithm to calculate uncollided photon fluence from a point source is straightforward. The computation-intensive portion of the discrete ordinates method is the sweep process that consumes 80-99% of the run time. In order to accelerate the discrete ordinates sweep process on a GPU, the concurrent tasks at any given moment must be known. In the single-threaded version, the solver sweeps through the voxels, one by one, in a pre-determined fashion based on the photon stream direction. At the very beginning, the input fluence to a single voxel is known from its boundary conditions. However, once that voxel's fluence is computed, all three of its outgoing fluence values enable three voxels to be computed independently of each other. After those three, six can be computed. Each layer of voxels whose fluence can be computed independently, is called a "sub-sweep." **Figure 1** illustrates some sub-sweeps through a cubic mesh.

In order to parallelize the sweep through the mesh, the global index of each voxel that can be computed in the $S$ sub-sweep must be known. When those indices are known, each voxel in sub-sweep $S$ can be solved independently by a GPU kernel. This task is greatly simplified by noticing that the $x$, $y$, and $z$ indices of all voxels in the fourth sub-sweep (shown in **Figure 1)** all sum to 4, as shown in **Table 1**.

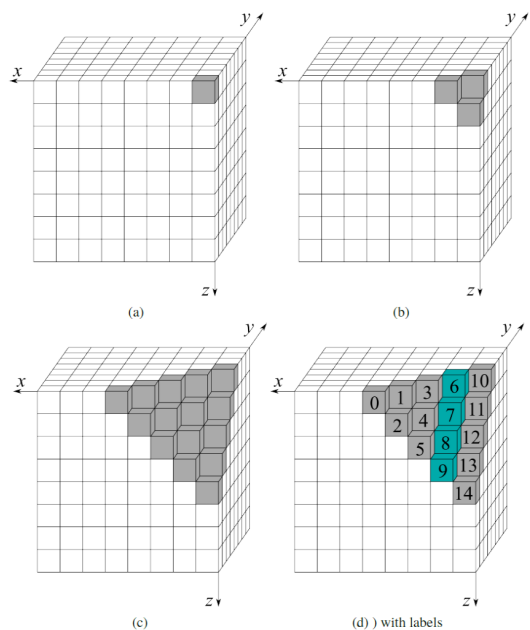

**Figure 1**. The progression of sub-sweeps throughout a sweep. Each sub-sweep must complete before those after it. Each voxel within a sub-sweep can be solved in parallel with all others in its sub-sweep. (a) Sub-sweep 0 ($S$ = 0); (b) Sub-sweep 1 ($S$ = 1); (c) Sub-sweep 4 ($S$ = 4); (d) Sub-sweep 4 ($S$ = 4) with labels.

The special case of the cube (shown in **Figure 1)** can be extended to a more general case, as illustrated in **Figure 2**. Notice that the aforementioned intuition that $i_x + i_y + i_z = S$ is not the sufficient and necessary condition. The final sub-sweep ($S$ = 14) contains a single voxel, even though many combinations of three integers will add to 14. The additional constraint is

$$i_x < N_x \; ; \; i_y < N_y \; ; \; i_z < N_z \tag{8}$$

The constraints listed in equation (8) can be used to compute the total number of parallel tasks in any sub-sweep. The number of parallel tasks, $P$, that can be done on sub-sweep $S$ of an $N_x \times N_y \times N_z$ mesh is given in equation (9), which is basically the number of voxels in the wave front moving through the spatial mesh grid (as shown in **Figure 2)**. Intuitively, the number of voxels in the wave front would be given by equation (10), if there were no boundaries in the three dimensions. With boundaries $N_x$, $N_y$, and $N_z$ in the three dimensions, the total number of voxels in the wave front could be corrected by equations (11)~(13). If the wave front passed the center of the mesh grid, the total number of voxels in the wave front could be corrected again by equations (14)~(16).

$$P = C_s - L_x - L_y - L_z + G_{xy} + G_{yz} + G_{xz} \tag{9}$$

where $C_s, L_x, L_y, L_z, G_{xy}, G_{yz}$ and $G_{xz}$ are defined by equations (10)-(22).

$$C_s = \frac{(S+1)(S+2)}{2} \tag{10}$$

$$L_x = \frac{d_x(d_x+1)}{2} \tag{11}$$

$$L_y = \frac{d_y(d_y+1)}{2} \tag{12}$$

$$L_z = \frac{d_z(d_z+1)}{2} \tag{13}$$

$$G_{xy} = \frac{d_{xy}(d_{xy}+1)}{2} \tag{14}$$

$$G_{yz} = \frac{d_{yz}(d_{yz}+1)}{2} \tag{15}$$

$$G_{xz} = \frac{d_{xz}(d_{xz}+1)}{2} \tag{16}$$

$$d_x = \max(S+1-N_x, 0) \tag{17}$$

$$d_y = \max(S+1-N_y, 0) \tag{18}$$

$$d_z = \max(S+1-N_z, 0) \tag{19}$$

$$d_{xy} = \max(S+1-N_x-N_y, 0) \tag{20}$$

$$d_{yz} = \max(S+1-N_y-N_z, 0) \tag{21}$$

$$d_{xz} = \max(S+1-N_x-N_z, 0) \tag{22}$$

Each voxel in a sub-sweep can be computed in parallel. Mathematically, the *i*th sub-sweep from all directions can be computed in parallel. However, in practice, this results in a race condition on the GPU hardware as the value of photon fluence in a voxel could be accessed/modified by different sweeps at the same time. As such, parallelization in an angular domain is not implemented in the current version of DOCTORS.

**Table 1**. Sub-sweep Indices

| $i$ | $i_x$ | $i_y$ | $i_z$ | $i_x + i_y + i_z$ |
|-----|-------|-------|-------|-------------------|
| 0   | 4     | 0     | 0     | 4                 |
| 1   | 3     | 1     | 0     | 4                 |
| 2   | 3     | 0     | 1     | 4                 |
| 3   | 2     | 2     | 0     | 4                 |
| 4   | 2     | 1     | 1     | 4                 |
| 5   | 2     | 0     | 2     | 4                 |
| 6   | 1     | 3     | 0     | 4                 |
| 7   | 1     | 2     | 1     | 4                 |
| 8   | 1     | 1     | 2     | 4                 |
| 9   | 1     | 0     | 3     | 4                 |
| 10  | 0     | 4     | 0     | 4                 |
| 11  | 0     | 3     | 1     | 4                 |
| 12  | 0     | 2     | 2     | 4                 |
| 13  | 0     | 1     | 3     | 4                 |
| 14  | 0     | 0     | 4     | 4                 |

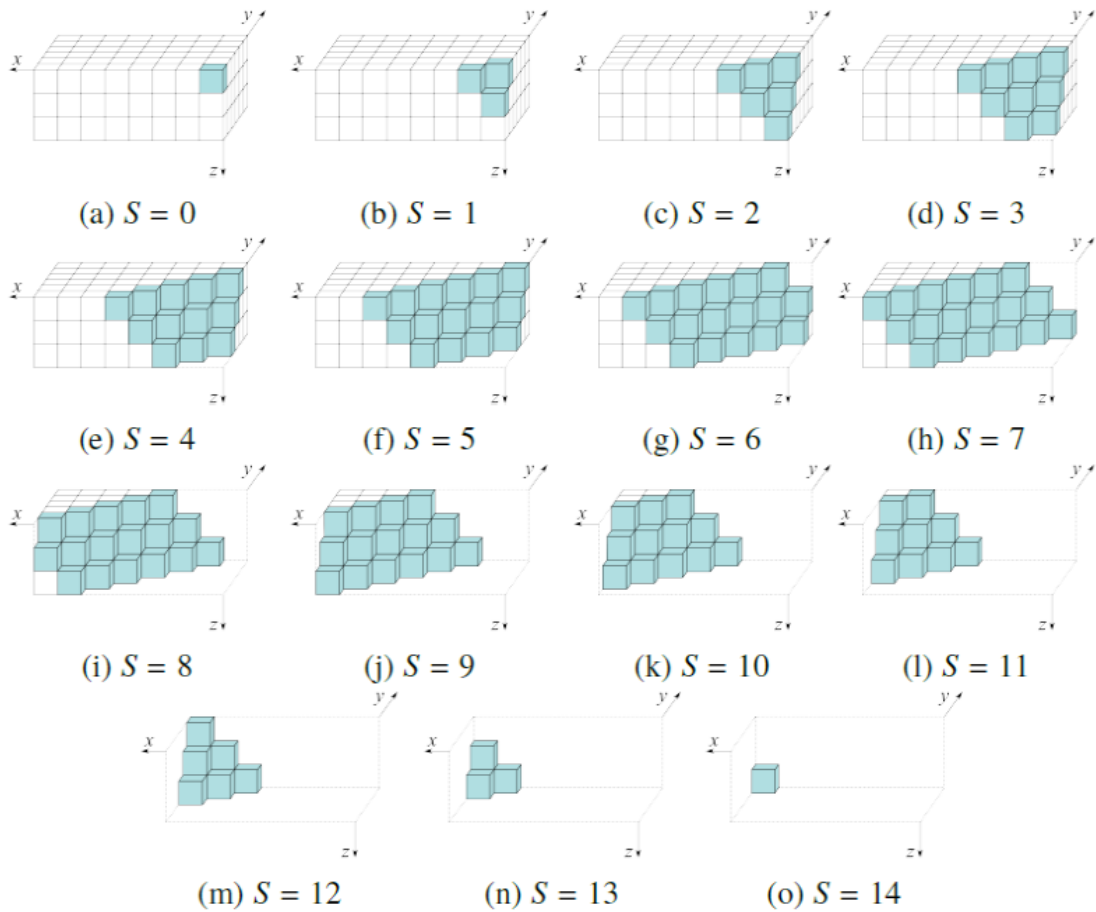

**Figure 2**. The generalized sub-sweep.

## II.C. The Computational Framework

DOCTORS was designed to be modular and its user interface was built using an open-source platform – Qt.  Qt is a framework for developing cross-platform application software. The major components of DOCTORS are shown in **Figure 3**. In the current version of DOCTORS, the *Geometry*, *Material*, and *Visualization* modules were created for users to input the object configuration. The *Multi-group Cross-section Library* module was created for users to select an appropriate pre-calculated multi-group photon cross-section library. The module of *User Input* was implemented for users to input specific configurations of the X-ray imaging system, such as source type, source spectrum, etc. The *Pre-Processing* module processes all of the information input into the DOCTORS and collects data for the 3D discrete ordinates solver. The 'ray-effect' is prominent in a CT imaging simulation when using the discrete ordinates method. To mitigate the 'ray-effect', the standard first-collision source method is employed in the *Ray Effects Mitigation* module. The 3D discrete ordinates solver solves the linear Boltzmann transport equation iteratively. The iterative solver (both CPU and GPU versions) are implemented in the current version of DOCTORS. An *Output File* module is a collection of the DOCTORS output, including photon fluence distribution, activity log of the computation, etc. The *Post-Processing* module is optional, where users can manipulate the results obtained from the solver, such as a conversion of photon fluence to an absorbed dose.

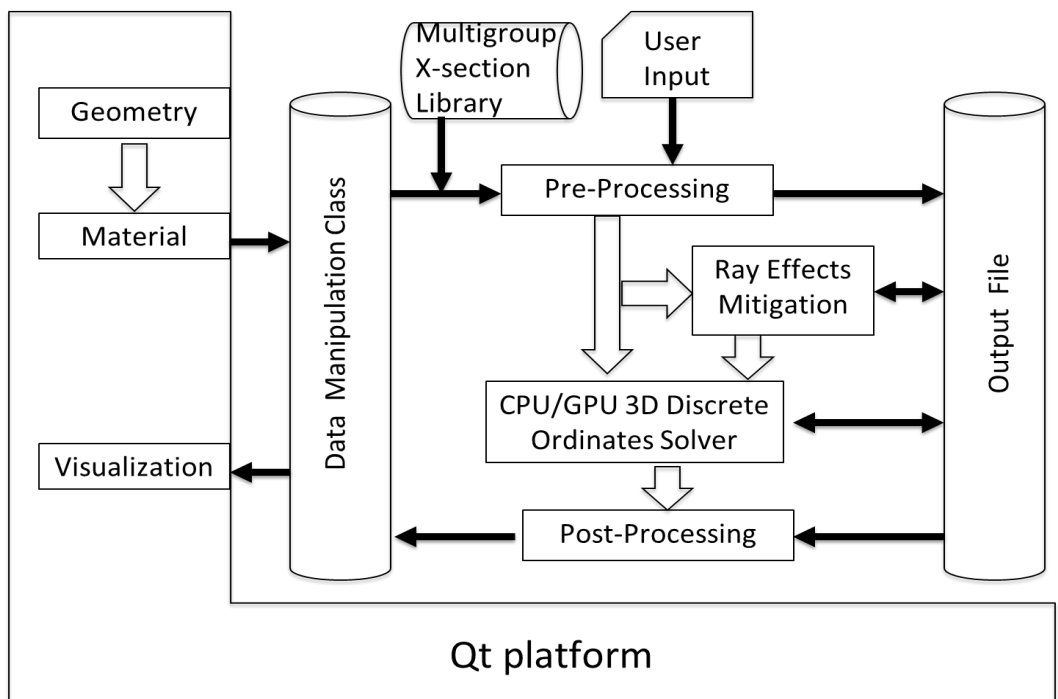

**Figure 3**. Modules of the computational framework.

The graphical user interface (GUI) of DOCTORS is shown in **Figure 4**. The first tab is the geometry input where a user provides 3-D reconstructed images, such as DICOM (digital imaging and communications in medicine) images or raw CT numbers. As soon as the data is read in, it is converted into a series of dosimetrically equivalent materials representative of a real object. The conversion between image pixel values to material type is automatic for medical applications. Although there is no direct relation between CT numbers and material types, CT numbers can be converted to material types quite accurately based on a stoichiometric calibration.[14-16]

Some screen shots of the geometry explorer are shown in **Figure 5**. The geometry viewer can show slices through the voxel phantom along all three major planes. The depth of a slice viewed is changed by moving the scroll bar in the center of the viewer up or down. As the scroll bar moves, the number at its bottom is updated, accordingly, to indicate the depth (in voxels) of the slice. In addition to the material, the geometry viewer can render the physical density (in g/cm$^3$) and the atom density (in atom/barn-cm), as well as the raw CT number used to generate the material and density.

The next tab is the cross-section dialog, which allows users to select and explore pre-calculated multi-group photon cross-sections. DOCTORS does not generate multi-group cross sections in situ from the ENDF/B or similar data library. Instead, the multi-group cross sections must be pre-calculated using either NJOY[17] or AMPX.[18] Currently, two types of multi-group cross-section library are supported in DOCTORS. The DTFR-type of library, generated by NJOY and the AMPX-type of library generated by AMPX. As an example, the 90-group cross sections of Oxygen are shown in **Figure 6**. The scattering matrix of the first order Legendre expansion is shown in **Figure 7**.

The next two tabs identify the quadrature, anisotropy treatment, and whether GPU is used for the photon transport computation. The final tab, as shown in **Figure 4,** defines the X-ray source. A user can select a source type from a number of built-in options available for analysis, including point sources, fan beams, and cone beams. Multi-fan beams and multi-cone beams can be arranged about an object to mimic the source rotation in a CT scan. Each source type is described by its position and energy distribution. Two additional parameters, $\varphi$ and $\theta$, describe the azimuthal and polar angles subtended by the fan or cone beam, respectively. It is worth noting that the source can be positioned anywhere in the geometry, including in the patient's organ. DOCTORS can be used to simulate gamma emission imaging techniques, such as SPECT (single photon emission computed tomography), or other complicated scan geometries.

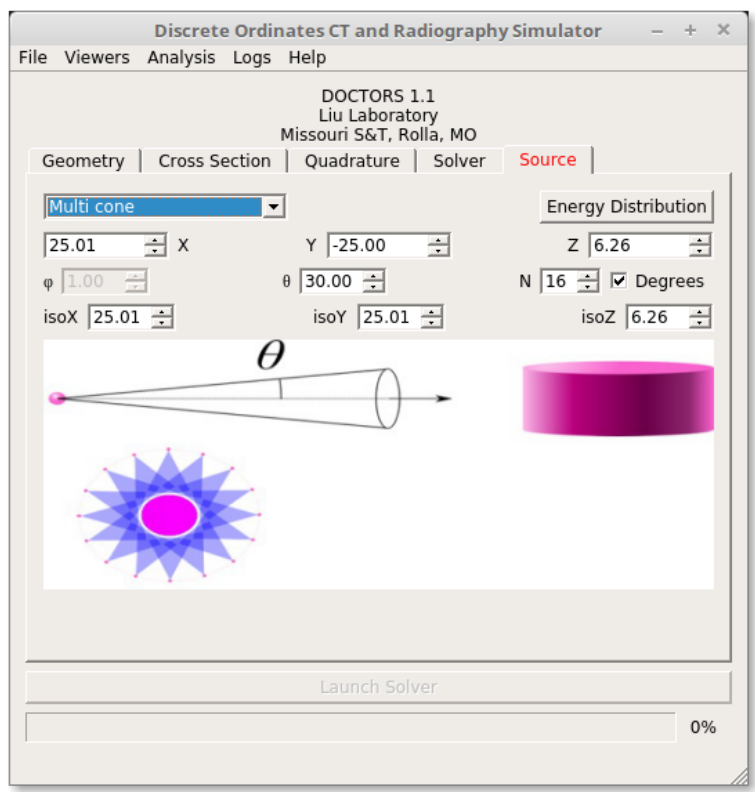

**Figure 4**. The graphical user interface of DOCTORS.

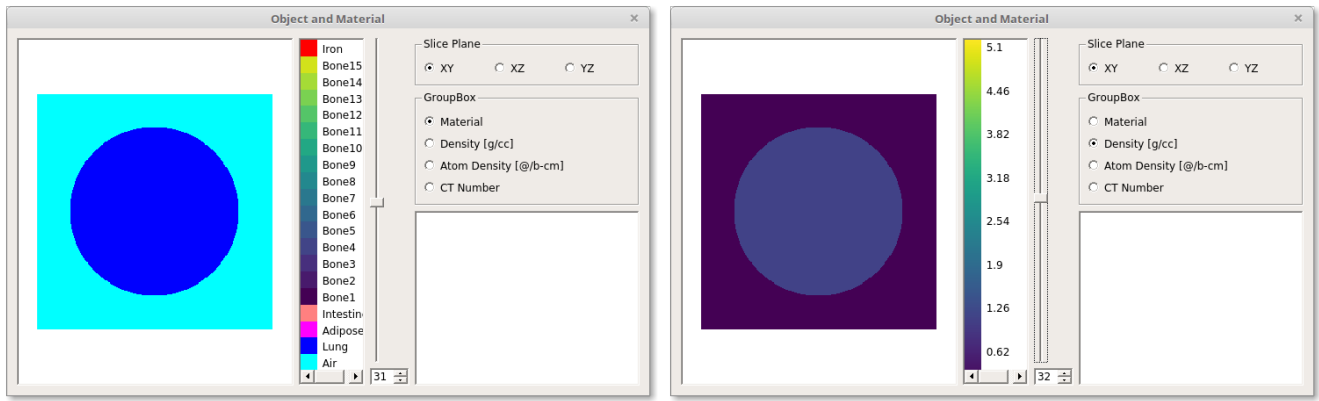

**Figure 5.** The geometry explorer of DOCTORS. Left: the material viewer which shows materials converted from a CT image of a water phantom; Right: the density of materials converted from the same CT image.

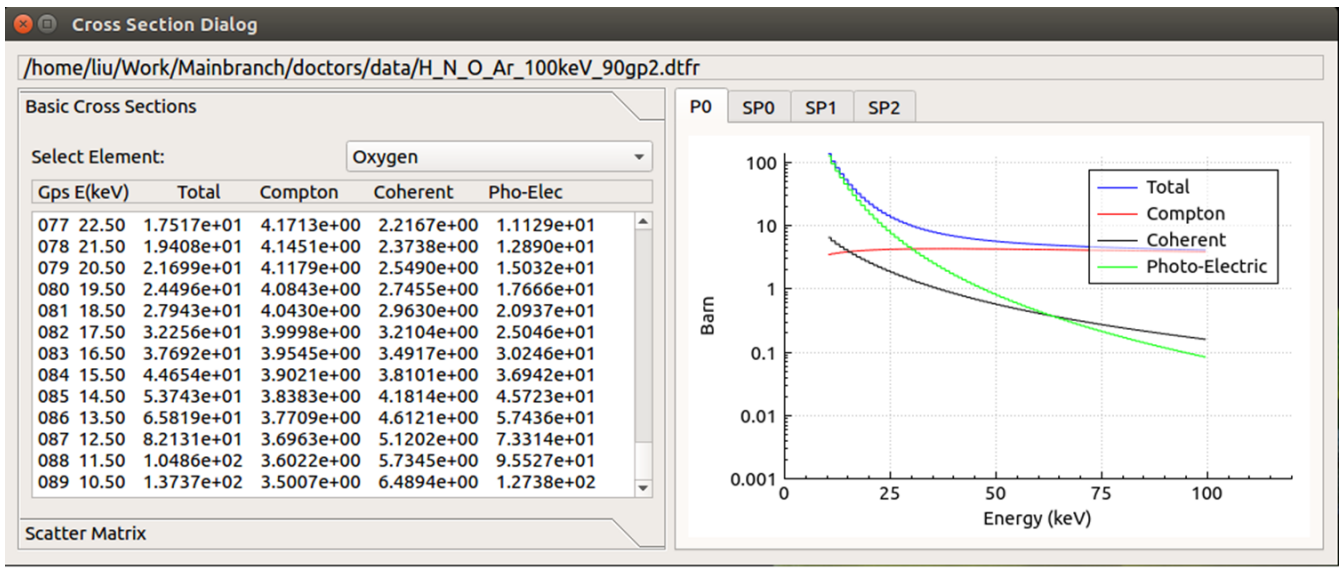

**Figure 6.** The cross section explore dialog of DOCTORS. An example shown in the dialog is the 90-group cross section of Oxygen, including total, photo-electric, Compton, and Coherent scattering cross sections.

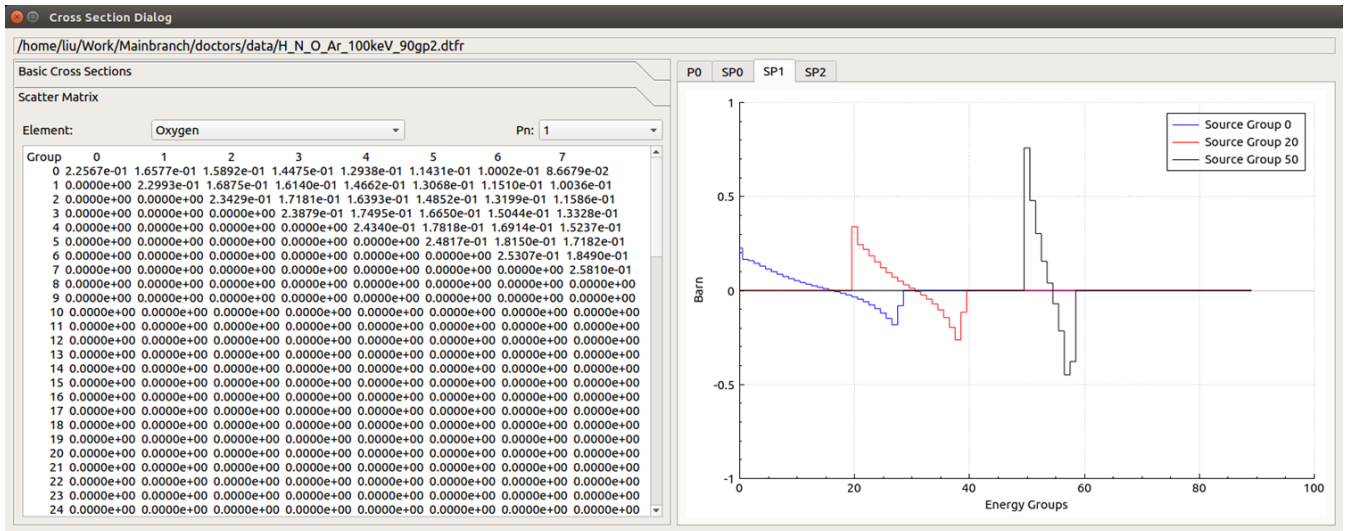

**Figure 7.** A scattering matrix cross section explore dialog of DOCTORS. The example shown in the dialog is the first order Legendre expansion of the group-to-group scattering cross section.

Once all necessary data is loaded, the "Launch Solver" button becomes active and will remain so unless the user creates a conflicting set of inputs that would prevent the solver from being able to run. When the user clicks on the "Launch Solver" button, the computation process begins and the photon fluence information will be displayed in the output dialog, as shown in **Figure** 8. The output dialog has controls very similar to those of the geometry explorer. Users can freely choose a plane to slice through the solution and can plot the solution on either a linear or logarithmic scale. The user can select whether the uncollided, collided, or total fluence is rendered. Since the linear Boltzmann transport equation is solved in an iterative fashion, the output dialog is progressively updated following each iteration so that the user can monitor the development of the solution.

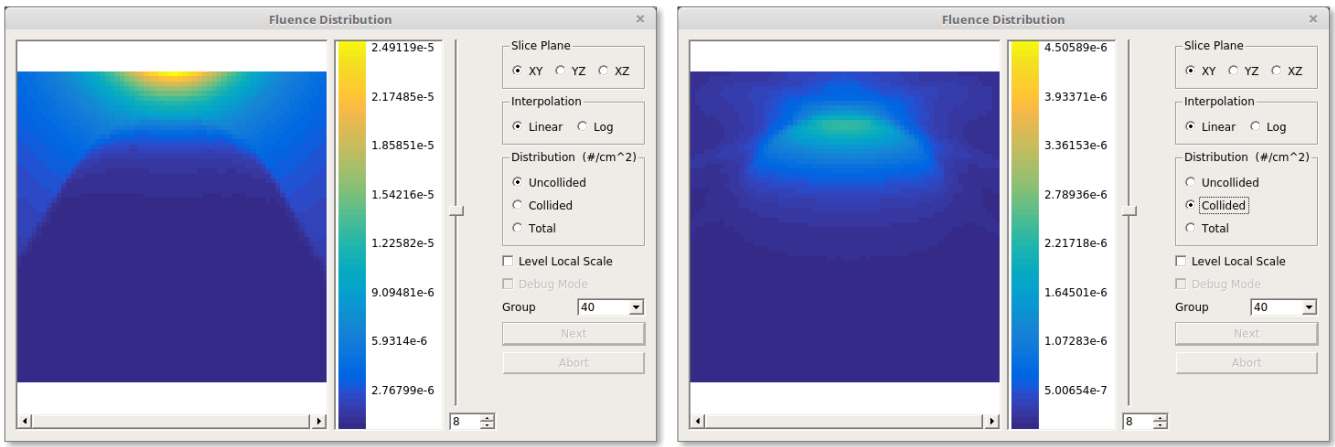

**Figure 8.** The output dialog of DOCTORS. Left: the uncollided photon fluence distribution of a single X-ray projection image. The X-ray source is located on the top of the water phantom; Right: the collided photon flux distribution of the same projection image.

## II.D. Verification against Monte Carlo Simulation

DOCTORS has the capability to automatically generate an input file for the Monte Carlo simulation, using geometry and material data and source specification provided by the user. The user can easily verify the results obtained from DOCTORS by comparing them to a Monte Carlo simulation with the exact same input parameters. This feature greatly reduces the amount of work required to manually generate a Monte Carlo simulation input file. Currently, the only supported file format for a Monte Carlo simulation is the MCNP[19] input file format. The version of MCNP used in this study is MCNP6. No variance reduction techniques were employed in Monte Carlo simulation in order to achieve a perfect analog to the photon transport in the object. To achieve good statistics (i.e., <5% uncertainty), $1 \times 10^9$ histories were used in all Monte Carlo simulations. To compare the DOCTORS and MCNP6 data, the root-mean-square-deviation (RMSD) was calculated using equation (23).

$$RMSD = \sqrt{\frac{\sum_{n=1}^{N}\left(\frac{DOCTORS_n - MCNP_n}{MCNP_n}\right)^2}{N}} \qquad (23)$$

The verification processes were divided into two categories: single projection (i.e. radiography) verification and CT scan verification. The single projection verification compared the calculated collided and uncollided photon fluence in each energy group for a single projection from a point source between DOCTORS and the MCNP6 simulations. The CT scan verification compared the same quantities of DOCTORS and MCNP6 for 16 concurrent projections placed uniformly around the phantom, which mimiced a CT scan.[6] All simulations were performed on a personal computer (PC), with an Intel i7-5960X processor and a base clock speed of 3.5 GHz.

Two source types were used in MCNP6. For the single projection verification, an isotropic point source was placed on top of the phantom. For the CT scan verification, 16 multi-cone beam sources were placed around the phantom to form a ring-shape structure and mimic the CT scan. The cone angle was 30° for each cone beam. A simple diagram about the singe cone-beam source and the ring-shape multiple cone-beam sources can be found in **Figure 4**. The source spectrum definition in MCNP6 was automatically imported from DOCTORS' user input. The cutoff energy of photon transport was 1 keV. Since the current version of DOCTORS does not simulate electron transport, electron transport was turned off in MCNP6 simulations. The cross section library used in MCNP6 was the photon library from ENDF/B VII.1. All other transport parameters were set to the default values. The 3D geometry and material type and density in each cell were automatically imported from DOCTORS. The tally type used was the photon fluence averaged over a cell (F4 tally). The photon fluence results were normalized to per source particle. Both collided and uncollided photon fluence were generated for each cell at specified energy bins.

## II.E. Deterministic Simulation Setup

A water cylinder phantom with a 35-cm diameter was used for verification of the DOCTORS code. The water phantom was located at the center of a rectangular space with dimensions of 50 cm x 50 cm x 12.5 cm in the x, y, and z directions, respectively. A simplified geometry of 64 x 64 x 16 meshes, with only 65,536 voxels (the voxel pitch is 7.81 mm), was used to speed up the Monte Carlo simulation. Theoretically, the mesh size should have had little impact on the accuracy of fluence calculation, as long as the mesh size was much smaller than the mean-free-path of photons. The mean-free-path of a photon, at 20 keV, was 13.9 mm in pure water, and monotonically increased to 60.6 mm at 100 keV. To benchmark DOCTORS' performance on a more realistic geometry, a finer mesh grid 256 x 256 x 64 with voxel pitch 1.95 mm of the same water phantom, was also used. In addition, to verify DOCTORS' performance on an inhomogeneous phantom, an abdomen phantom was used in the CT scan verification. The abdomen phantom had the same dimensions (50 cm x 50 cm x 12.5 cm) as the water phantom in the x, y, and z directions. The cross section view of the abdomen phantom is shown in **Figure 9.**

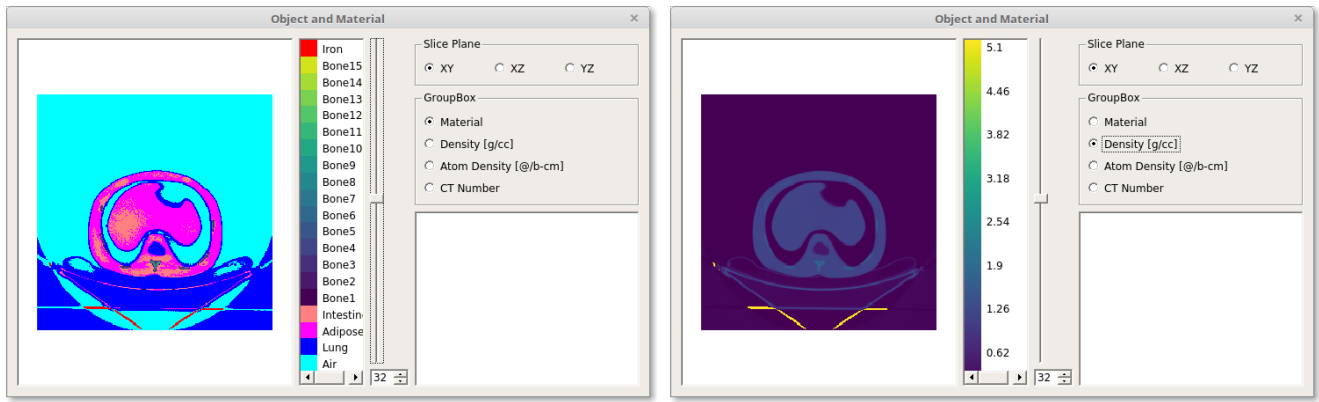

**Figure 9.** The cross-section view of an abdomen phantom used in DOCTORS verification. Left: the material viewer shows materials converted from a CT image; Right: the density of materials converted from the same CT image.

In DOCTORS, four different quadrature sets $S_2$, $S_4$, $S_6$, and $S_8$ could be selected, which represented 8, 24, 48, and 80 directions, respectively. The order of Legendre polynomial expansion varied from 0 to 2. The peak energy of the X-ray used in the simulations was 100 keV, and 90 photon energy groups, with 1 keV interval from 10 keV to 100 keV, were used. **Figure 10** shows the X-ray spectrum and the 90 groups. The 90-group cross sections used in DOCTORS were created by the NJOY program using data from the ENDF/B-VII.0 cross section library.[20]

As indicated by our previous study,[10] isotropic scattering is dominant at such low energy photons (<100 keV). Therefore, quadrature set $S_6$ and isotropic scattering (i.e., zeroth order Legendre expansion) were used in all simulations.

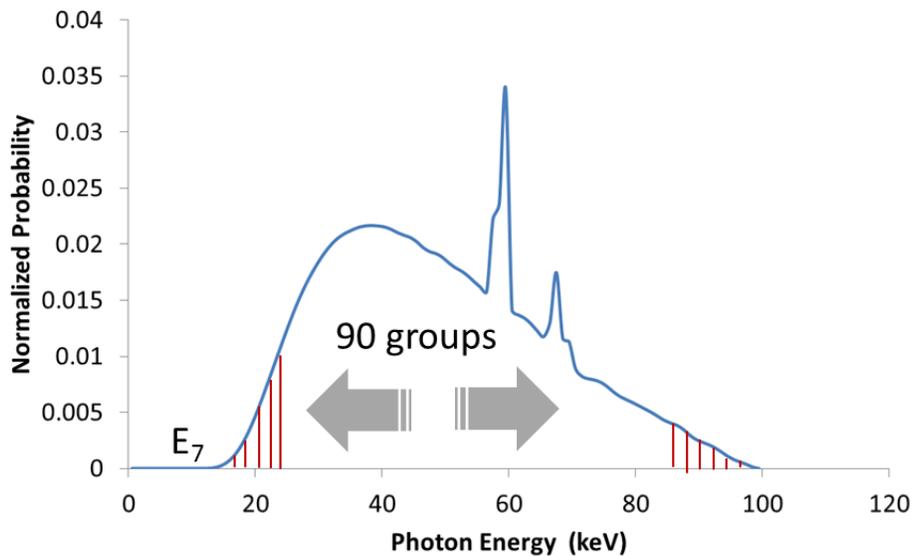

**Figure 10**. The X-ray spectrum and 90 energy groups used in DOCTORS.

## II.E. GPU Acceleration

In the current version of DOCTORS, two kernels including parallel ray-tracing and voxel sweeping algorithms were implemented on a single GPU architecture, using CUDA[21] language.

The GPU card used was NVidia GeForce GTX Titan-z, which actually had two GPUs on one card. Each GPU had 6 GB memory. Since one of the GPUs is mainly used for display purposes, only one GPU was used for computation in the current implementation of DOCTORS. The most commonly used global memory was adopted in the current GPU implementations. The performance of GPU acceleration was benchmarked for both X-ray single projections and CT scans of the water phantom.

## III. Results

### III.A. Single Projection Verification

Single projections from an isotropic point source on top of the phantom were simulated using both MCNP6 and DOCTORS. The uncollided and collided photon fluence distributions in each energy group of the central slice, calculated by the two different methodologies, were compared using RMSD. The RMSD results are plotted in **Figure 11**. An example of cross-section views of the collided and uncollided photon fluence distributions, as well as the difference map between DOCTORS and MCNP6 simulations, are shown in **Figure 12** and **Figure 13**, respectively. The computer run time of DOCTORS was 4.13 minutes. A large number of histories was necessary for MCNP6 to achieve <5% one standard uncertainty in most of voxels. The average of one standard uncertainty was 1.7% and 2.4% for the uncollided and collided photon fluence estimations, respectively. Nevertheless, the photon fluence simulated by MCNP6 in the energy ranges of 10 keV to 20 keV and 90 keV to 100 keV were discarded due to high statistical uncertainties.

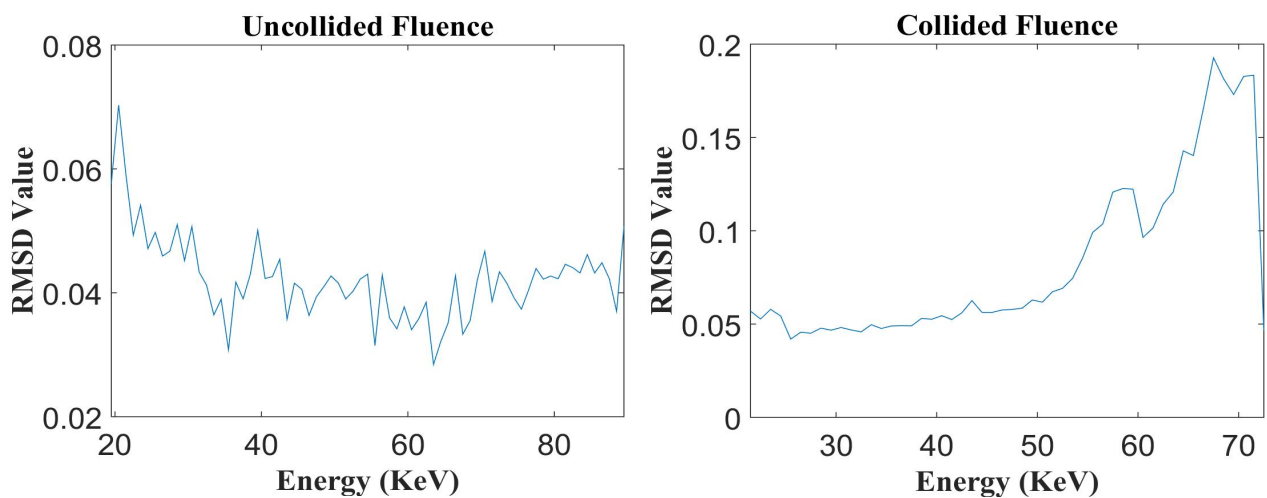

**Figure 11.** Comparison of RMSD values of uncollided and collided photon fluences from a single projection in a water phantom. RMSD values were only calculated when the uncertainty of the MCNP estimation was less than 5%.

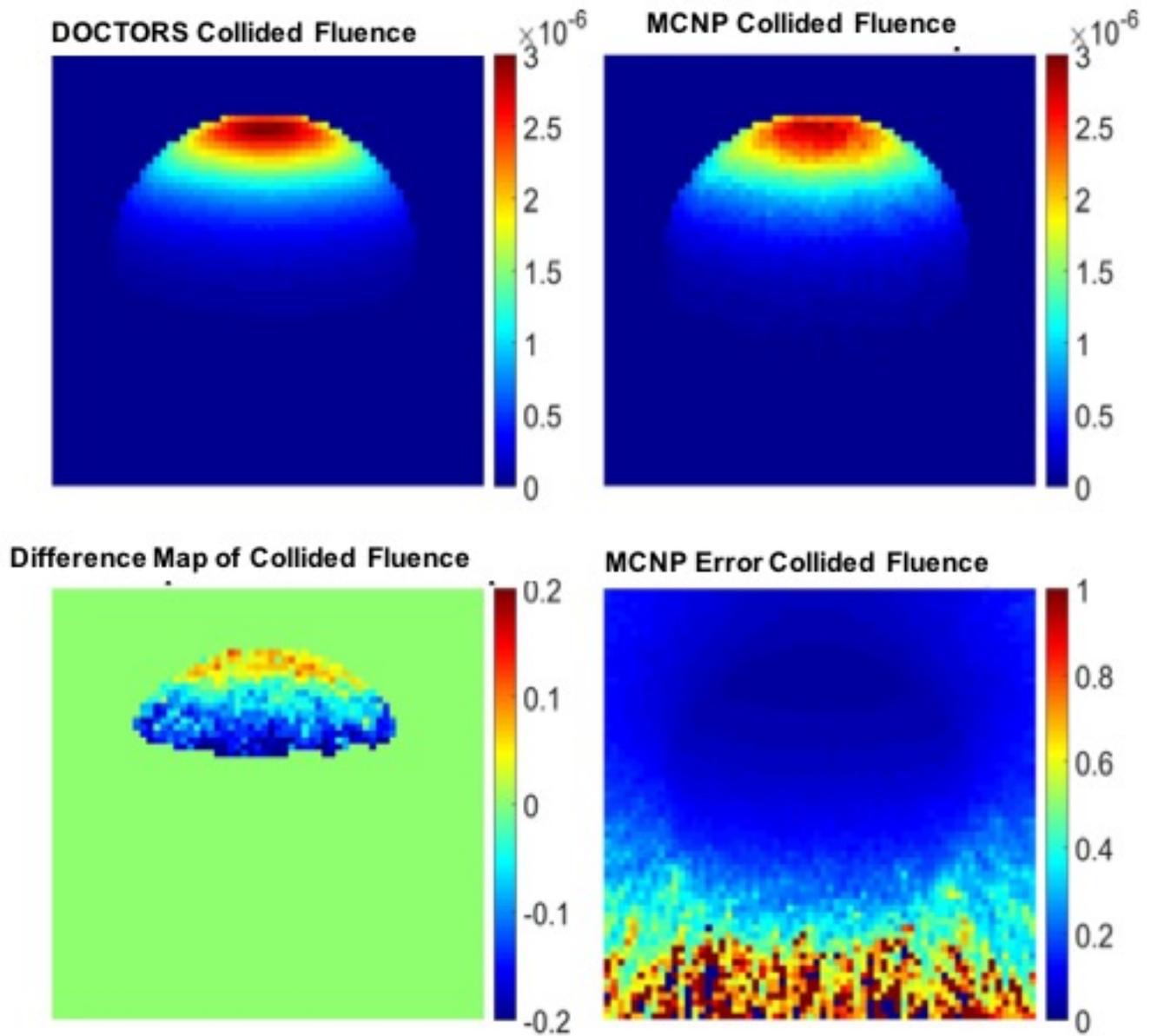

**Figure 12.** Cross-section view of collided photon fluence distribution (photons/cm$^2$ per source particle) from a single projection of a water phantom at energy level 54~55 keV, calculated by DOCTORS and MCNP6. The difference map and the MCNP error map are shown in absolute values.

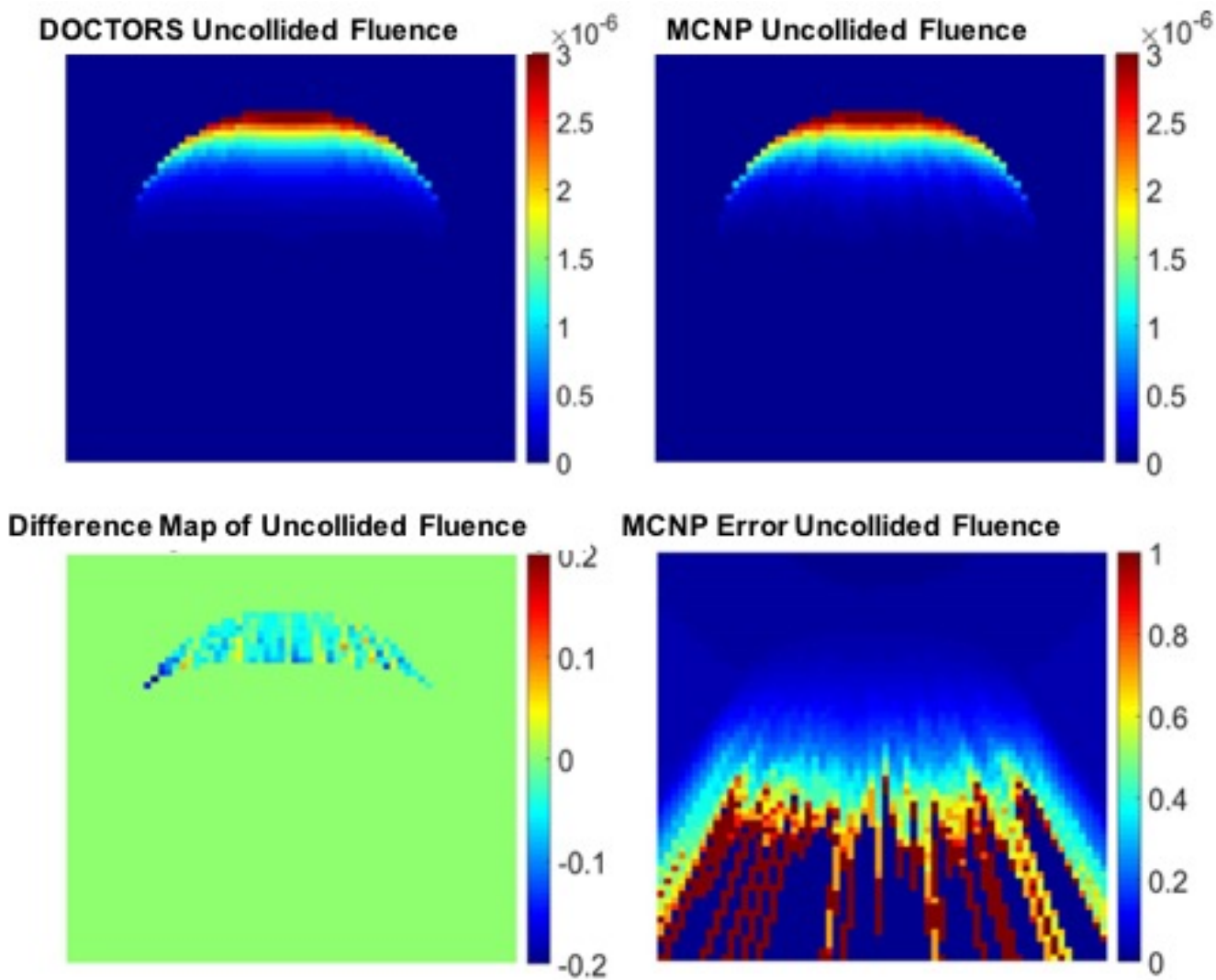

**Figure 13.** Cross-section view of uncollided photon fluence distribution (photons/cm$^2$ per source particle) from a single projection of a water phantom at energy level 54~55 keV, calculated by DOCTORS and MCNP6. The difference map and the MCNP error map are shown in absolute values.

## III.B. CT Scan Verification

The CT scan simulation was realized by uniformly placing 16 cone beam sources around the water phantom. The RMSD results are plotted in **Figure. 14**. An example cross-section view of the collided and uncollided photon fluence distribution, as well as the difference map between DOCTORS and MCNP6 simulations are shown in **Figure 15** and **Figure 16**, respectively. It only took 5.08 minutes for DOCTORS with the quadrature set $S_6P_0$.

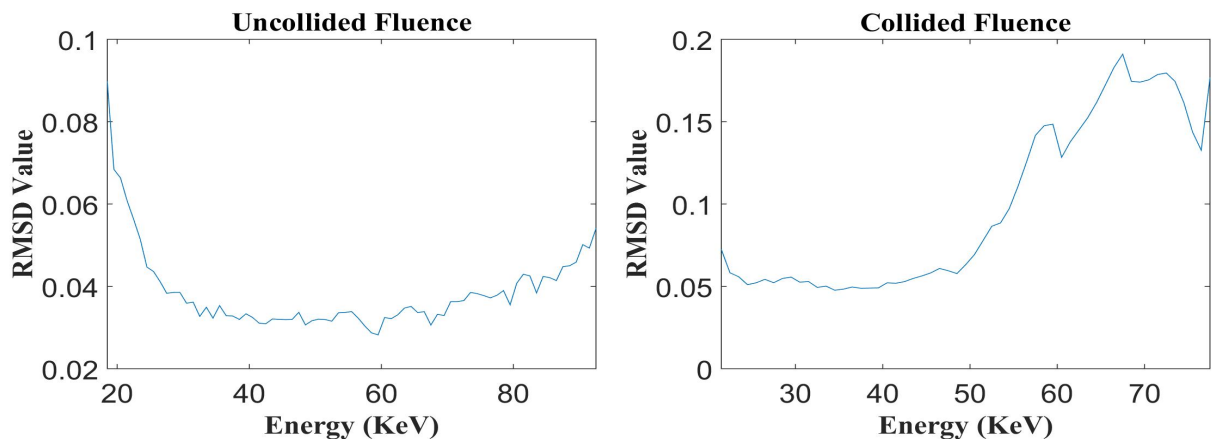

**Figure 14.** Cross-section view of energy-resolved collided photon fluence distribution (photons/cm$^2$ per source particle) from a CT scan of a water phantom, calculated by MCNP6 and DOCTORS with quadrature set $S_6P_0$.

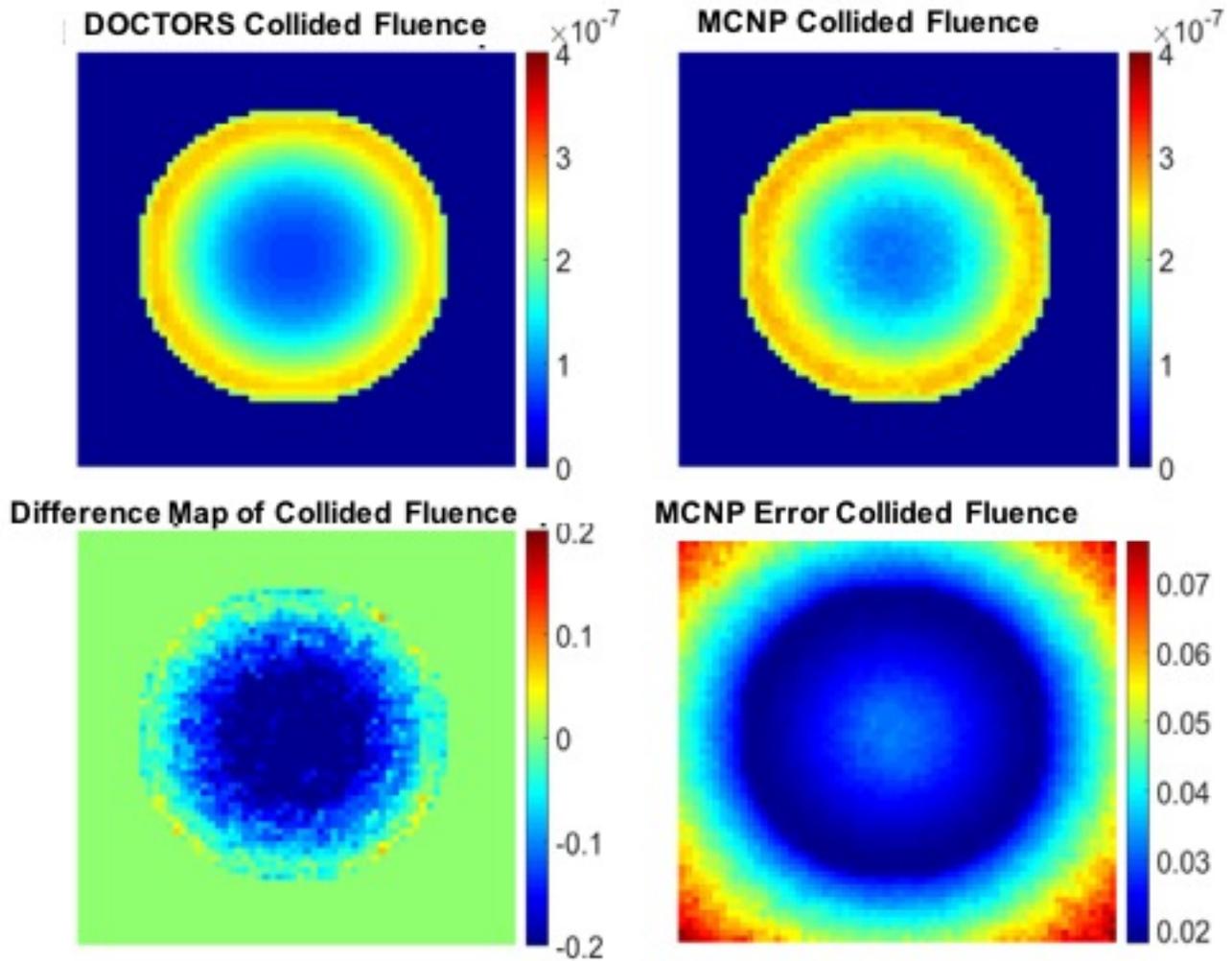

**Figure 15.** Cross-section view of collided photon fluence distribution (photons/cm$^2$ per source particle) from a CT scan of a water phantom, at energy level 54~55 keV, calculated by DOCTORS and MCNP6. The difference map and the MCNP error map are shown in absolute values.

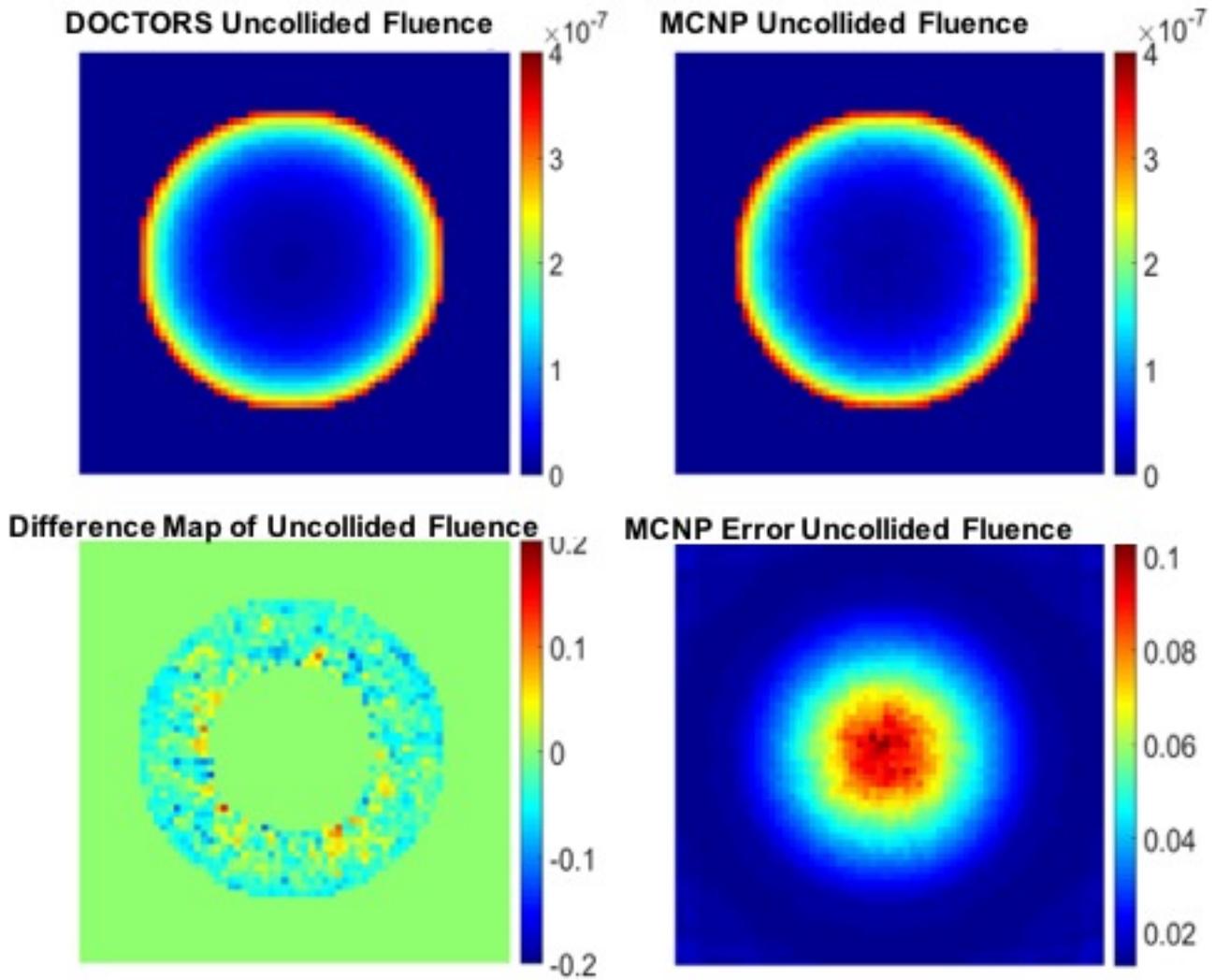

**Figure 16.** Cross-section view of uncollided photon fluence distribution (photons/cm$^2$ per source particle) from a CT scan of a water phantom, at energy level 54~55 keV, calculated by DOCTORS and MCNP6. The difference map and the MCNP error map are shown in absolute values.

The CT scan simulation was repeated on the inhomogeneous abdomen phantom using the same X-ray source setup and mesh grid. Examples cross-section views of the collided and uncollided photon fluence distributions, as well as a map of difference between DOCTORS and MCNP6 simulations, are shown in **Figure 17** and **Figure 18**, respectively.

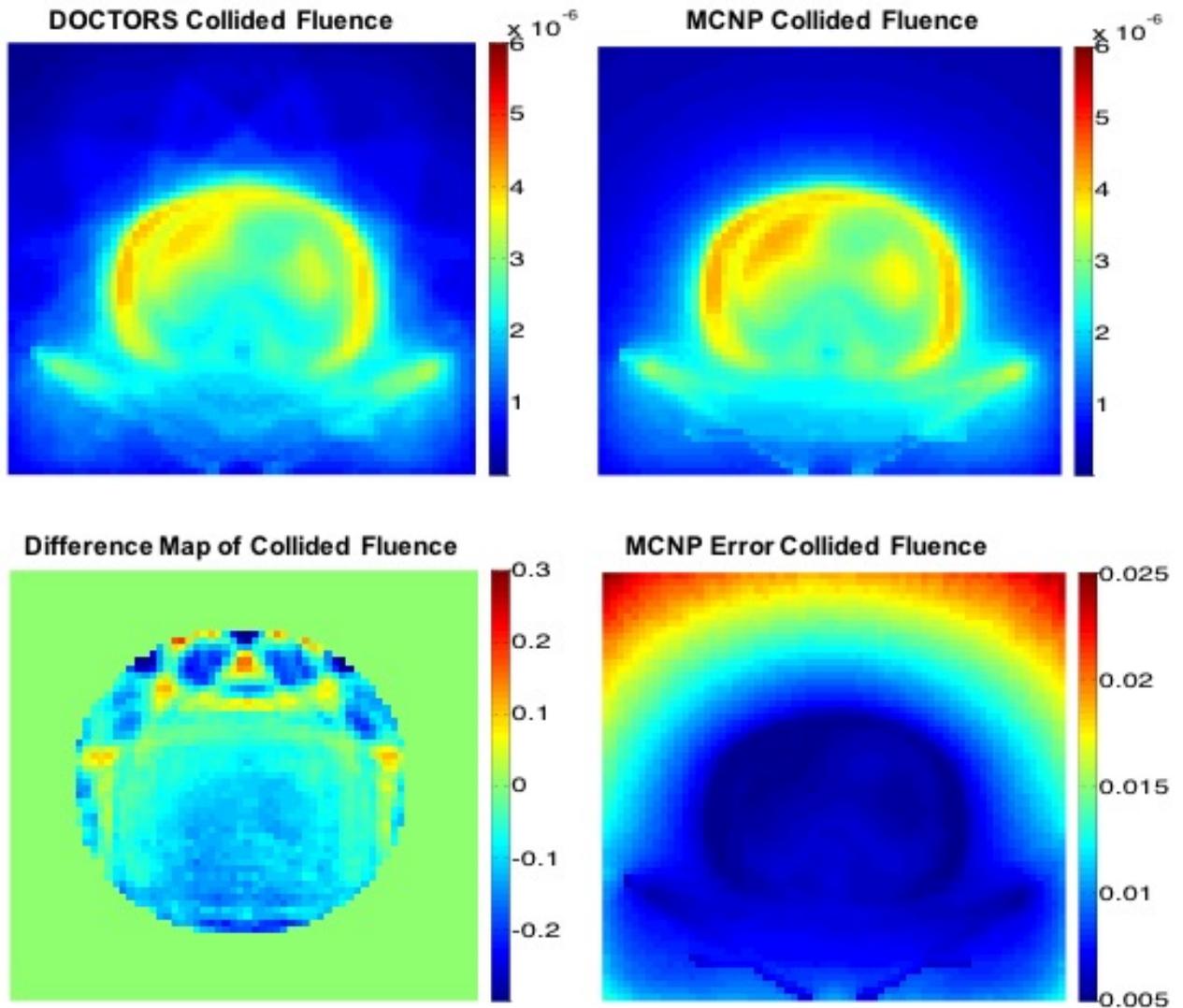

**Figure 17.** Cross-section view of collided photon fluence distribution (photons/cm$^2$ per source particle) from a CT scan of an abdomen phantom at energy level 30~45 keV, calculated by DOCTORS and MCNP6. The difference map and the MCNP error map are shown in absolute values.

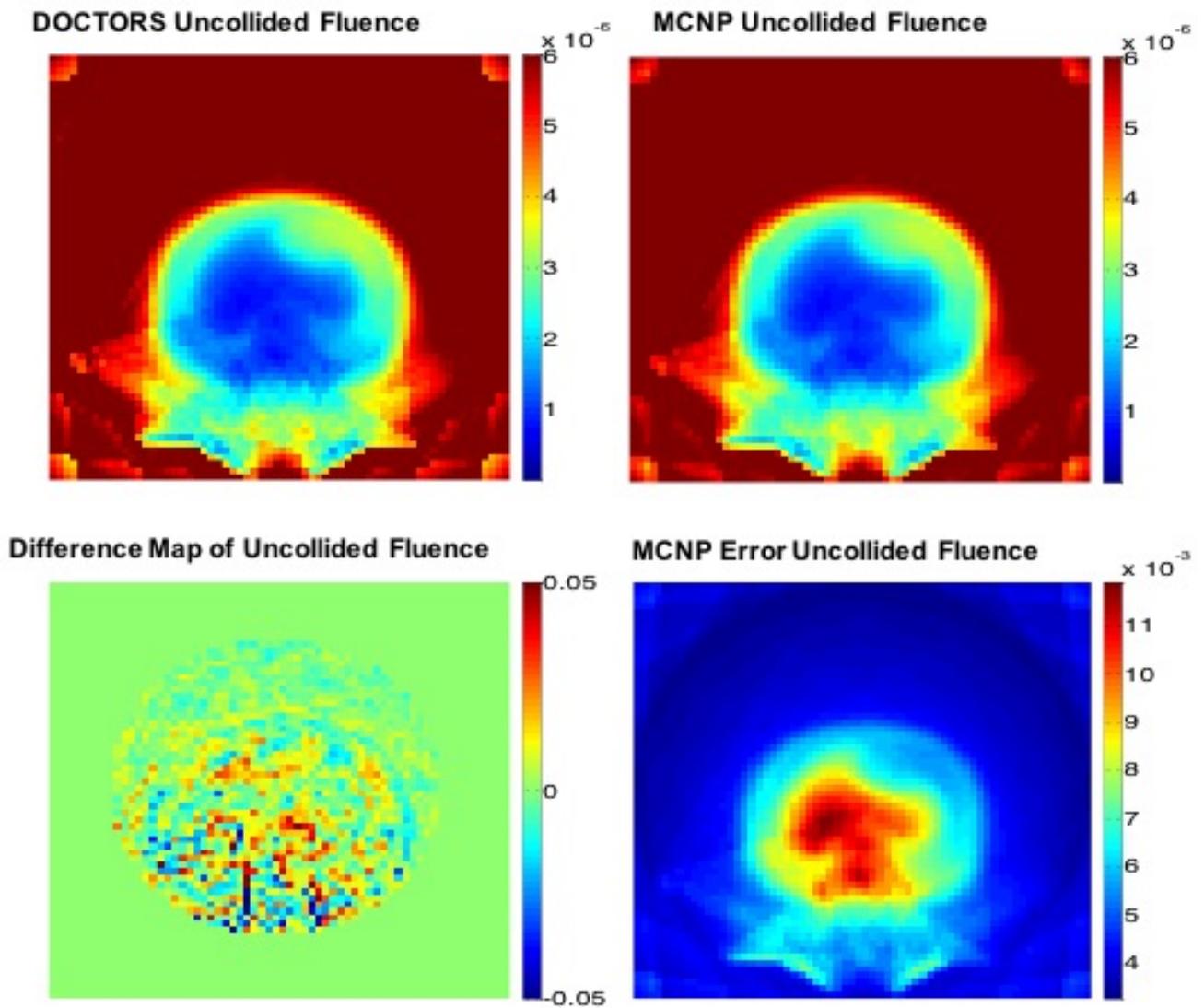

**Figure 18.** Cross-section view of uncollided photon fluence distribution (photons/cm$^2$ per source particle) from a CT scan of an abdomen phantom at energy level 30~45 keV, calculated by DOCTORS and MCNP6. The difference map and the MCNP error map are shown in absolute values.

### III.C. GPU Acceleration

Parallel ray-tracing and 3D voxel sweeping algorithms for X-ray single projections and CT scans were implemented on GPU architectures in DOCTORS. The GPU results were nearly identical to the results obtained from CPU. The maximum difference was between $\pm 0.02\%$. **Table 2** summarizes the runtime required for the CPU-only version of DOCTORS and the GPU accelerated version. The quadrature set used in DOCTORS was $S_6P_0$. The speedup of the GPU over the CPU is also given in **Table 2**.

**Table 2**. GPU runtime benchmark on coarse (64x64x16) and fine mesh (256x256x64) grids

| Time ╲ Mesh | Water Phantom | | | |
|---|---|---|---|---|
| | Coarse Mesh Single Projection | Coarse Mesh CT Scan | Fine Mesh Single Projection | Fine Mesh CT Scan |
| CPU (minutes) | 4.13 | 5.08 | 530.7 | 769.6 |
| GPU (minutes) | 0.71 | 0.72 | 16.9 | 81.9 |
| Speedup | 5.81 | 7.06 | 31.4 | 9.4 |

## IV. Discussion

As shown by the simulation results, both single projections and CT scans, the photon fluence distribution calculated by DOCTORS was in good agreement with MCNP simulations. The maximum RMSD value for uncollided photon estimation was below 10%, and the maximum RMSD value for collided photon estimation was below 20%. For collided photo fluence estimation, the RMSD value was below 10% if photon energy was less then 55 keV. The RMSD value is then increased as the photon energy increased. The uncollided photon fluence was calculated analytically in DOCTORS, so there were only very small discrepancies versus the MCNP simulations. Relatively large discrepancies were expected for collided photon estimations due to the fundamentally different treatment of scattered photons in DOCTORS and MCNP6. In medical diagnostic imaging using X-rays, the primary photon fluence (i.e., uncollided photon fluence) is dominant at higher X-ray energies, so the majority of photons are the uncollided photons in the total photon fluence, which were used in the calculation of absorbed dose. Therefore, any error in collided photon fluence had less impact on the final dose calculation. Nevertheless, the relatively large error in collided photon estimation is one of the limitations of DOCTORS.

As shown on the difference maps in Figures 11 and 14, DOCTORS overestimated the collided photon fluence at the boundary and underestimated the collided photon fluence near the center of the phantom. This can be explained by the spatial truncation error introduced by the diamond difference which was used in DOCTORS. To minimize the spatial truncation error, one solution is to decrease the mesh size. However, increasing the number of mesh points with smaller mesh size significantly increases calculation time.

As shown in Table 2, the computation time of DOCTORS was less than 5 minutes for a single projection in a physical volume of 50cm x 50cm x 12.5cm on an ordinary personal computer with a coarse mesh grid. This computation time can be further decreased, to less than 1 minute, by employing a parallel computing technique using GPU. For an axial CT scan, the computation time is still around 5 minutes on a personal computer, using a single CPU, and the computation time is, again, less than 1 minute using GPU parallel computing. This indicates that photon fluence and radiation dose distribution can be calculated nearly in real-time using GPU acceleration. On the other hand, the computation time was increased by a significant amount when a fine mesh grid was used as shown in Table 2. The tradeoff between the computation time and mesh grid size is an important factor to consider when using deterministic methods. It should be pointed out that the parallelization can be certainly implemented using multi CPUs or a CPU with multicores. If parallelization were implemented in the CPU version, the difference in performance between the GPU and CPU would be much smaller. The benefit of using GPU may not be significant when the problem size is small, such as the case of 64x64x16 mesh grid. However, for a larger problem (e.g., 256x256x64 mesh grid), the benefit using GPU maybe significant considering the number of cores available in a GPU (see Table 2).

As shown in figures 15~18, DOCTORS is capable of estimating photon fluence in CT scans of homogenous or inhomogeneous objects. However, it should be noted that the current version of DOCTORS does not simulate electron transport nor dose estimate photon generation from Bremsstrahlung radiation. Therefore, DOCTORS should be used with caution when the scanning object is made of heavy and thick materials.

## V. Conclusion

We have developed a computational framework for photon transport simulation in X-ray radiography and tomography using the discrete ordinates method. The accuracy of the discrete ordinates method was very close to that of a Monte Carlo simulation. Energy-resolved photon fluence from either a cone-beam single projection (i.e., radiography) or an axial scan (i.e., tomography) can be calculated within one minute.

**Acknowledgements**

This work is supported by the NRC Graduate Fellowship Grant (Grant Number NRC-HQ-13-G-38-0026), from the U.S. Nuclear Regulatory Commission.